\documentclass{report}
\usepackage{amsmath}

\newtheorem{theorem}{Theorem}
\newtheorem{acknowledgement}[theorem]{Acknowledgement}

\input{tcilatex}

\begin{document}

\title{Fermi Surface Renormalization in Two Spatial Dimensions}
\author{A. Ferraz \\
%EndAName
Laboratorio de Supercondutividade,\\
Centro Intl.de Fis.da Materia Condensada-UnB}
\maketitle

\begin{abstract}
We discuss the renormalization induced by interactions of a two-dimensional
truncated Fermi surface $\left( FS\right) $ model.Using a field theoretical
renormalization group method we calculate the critical renormalized physical
chemical potential $\mu _{R}$. We show that it either vanishes or approaches
a non-zero value. We argue that the vanishing of $\mu _{R}$ is indicative of
a further truncation of the $FS$ we started with and might well represent an
insulating spin liquid phase.
\end{abstract}

Strongly interacting fermion systems continue to attract great interest.
This was mainly motivated by the appearance of the high temperature
superconductors \ and by the realization that these materials have, at low
and optimal dopings, many anomalous properties. Besides, at zero doping they
are antiferromagnetic Mott insulators\cite{Anderson} and it is therefore
clear that strong correlation effects play a decisive role in this
phenomenon.

Metallic systems are readily characterized by their Fermi Surface$\left(
FS\right) $ and by their low-lying fermionic excitations. However in one
spatial dimension$\left( d=1\right) $, in the framework of a
Tomonaga-Luttinger liquid, the $FS$ is well defined but the low- lying
excited states are bosonic modes\cite{Emery},\cite{Voit}. The quasiparticles
are therefore destroyed by the interactions in $d=1$. Nevertheless the
non-interacting Fermi momentum $k_{F}$ continues to be a relevant physical
parameter and it is non-renormalizable not to violate Luttinger%
%TCIMACRO{\U{b4}}%
%BeginExpansion
\'{}%
%EndExpansion
s theorem \cite{Bedell} \cite{Affleck}. Thus in the Tomonaga-Luttinger
liquid particle-hole excitations continue to exist at $2k_{F}$ and the
interacting single-particle Green%
%TCIMACRO{\U{b4}}%
%BeginExpansion
\'{}%
%EndExpansion
s function is such that $\func{Im}G$ shows a change of sign at the
unrenormalized $k_{F}$.

We want to consider in this work what should happen to the $FS$ of strongly
interacting fermions in $d=2$. The importance of performing a proper $FS$
renormalization was also emphasized recently by Kopietz and Busche\cite%
{Kopietz} who treat the Fermi surface as a renormalization group$\left(
RG\right) $ fixed point manifold. Earlier on the the renormalization of the
Fermi surface for $2d$ electrons near a van Hove singularity was discussed
by Gonzalez, Guinea and Vozmediano\cite{Gonzalez} \ Here we approach this
problem using a field theoretical $RG$ method which allows us to deal with
these questions in a direct way. In higher dimensions the renormalization of 
$FS$ does not necessarily imply the violation of Luttinger%
%TCIMACRO{\U{b4}}%
%BeginExpansion
\'{}%
%EndExpansion
s theorem\cite{Luttinger}.

To initiate our discussion we consider one-particle fermionic states
described by the renormalized lagrangian density

\begin{equation*}
\emph{L}=Z\left( x\right) \psi _{\sigma }^{\dagger }\left( x\right) \left[
i\partial _{t}+\frac{\nabla ^{2}}{2}+\mu _{0}\left( x\right) \right] \psi
_{\sigma }\left( x\right) +
\end{equation*}

\begin{equation}
-\int_{\overline{x}}U_{0}\left( x-\overline{x}\right) Z\left( x\right)
Z\left( \overline{x}\right) \psi _{\uparrow }^{\dagger }\left( x\right) \psi
_{\downarrow }^{\dagger }\left( \overline{x}\right) \psi _{\downarrow
}\left( \overline{x}\right) \psi _{\uparrow }\left( x\right) 
\end{equation}

where $\mu _{0}$ is the bare chemical potential, $U_{0}$ is the bare
coupling function and $Z\left( x\right) $ is a local charge renormalization
parameter. Here the word bare is used in a field theory context and does not
mean non-interacting. Since \emph{L} is exact the Green%
%TCIMACRO{\U{b4}}%
%BeginExpansion
\'{}%
%EndExpansion
s function which result from it are already renormalized. The spatial
dependence of both $U_{0\text{ \ \ }}$and $Z$ in our model reflect the fact
that we use a truncated $2d$ Fermi surface which consists of four
symmetrical patches centred around $\left( \pm k_{F},0\right) $ and $\left(
0,\pm k_{F}\right) $ in $\mathbf{k}$-space as our unrenormalized $FS$. As we
show in Figure $1$ inside each patch in our model there are flat sectors in
the border regions which are continuously connected to a curved arc around
its center\cite{Ferraz}. This model was used before to describe
two-dimensional quasiparticles states sandwiched by one-particle states with
one-dimensional energy dispersion law to simulate the ``cold'' and ``hot''
spots\cite{Ferraz2}\cite{Hlubina} respectively. As a result of the momentum
space anisotropy the bare parameters become local functions of momenta and
the regularization is no longer isotropic and simple.

Following renormalization theory\cite{Collins} let us rewrite our lagrangian
density in a more convenient form:

\begin{equation*}
\emph{L}=\psi _{\sigma }^{\dagger }\left( i\partial _{t}+\frac{\nabla ^{2}}{2%
}+\mu \left( \omega \right) \right) \psi _{\sigma }-\int_{\overline{x}%
}U\left( x-\overline{x}\right) \psi _{\uparrow }^{\dagger }\left( x\right)
\psi _{\downarrow }^{\dagger }\left( \overline{x}\right) \psi _{\downarrow
}\left( \overline{x}\right) \psi _{\uparrow }\left( x\right) 
\end{equation*}

\begin{equation*}
+\psi _{\sigma }^{\dagger }\left[ \Delta Z\left( x\right) \left( i\partial
_{t}+\frac{\nabla ^{2}}{2}\right) +\Delta \mu _{0}\left( x\right) \right]
\psi _{\sigma }+
\end{equation*}

\begin{equation}
-\int_{\overline{x}}\Delta U_{0}\left( x-\overline{x}\right) \psi _{\uparrow
}^{\dagger }\left( x\right) \psi _{\downarrow }^{\dagger }\left( \overline{x}%
\right) \psi _{\downarrow }\left( \overline{x}\right) \psi _{\uparrow
}\left( x\right) 
\end{equation}

where $\mu \left( \omega \right) =\frac{1}{2}k_{F}^{2}\left( \omega \right) $
is the renormalized chemical potential parameter and $\omega $ is an energy
scale parameter. Besides, we have that

\begin{equation}
\Delta \mu _{0}\left( x\right) =Z\left( x\right) \mu _{0}\left( x\right) -%
\frac{1}{2}k_{F}^{2}\left( \omega \right) ,
\end{equation}
\begin{equation}
\Delta Z\left( x\right) =Z\left( x\right) -1
\end{equation}
and finally

\begin{equation}
\Delta U_{0}\left( x-\overline{x}\right) =U_{0}\left( x-\overline{x}\right)
Z\left( x\right) Z\left( \overline{x}\right) -U\left( x-\overline{x}\right)
\end{equation}
Here $\mu \left( \omega \right) $ is not directly associated with the
physical Fermi momentum which is associated with the number density. In fact
as we will see later $\mu \left( \omega \right) \rightarrow 0$ as we let $%
\omega \rightarrow 0$. Let us now use \emph{L} to calculate perturbatively
the renormalized one-particle irreducible function $\Gamma _{R}^{\left(
2\right) }\left( \mathbf{p},p_{0};\omega \right) $ for $\mathbf{p}$ in the
vicinity of the $FS$ point $\left( \Delta \left( \omega \right)
,-k_{F}\left( \omega \right) +\frac{\Delta ^{2}\left( \omega \right) }{%
2k_{F}\left( \omega \right) }\right) $. Up to order $O\left( U^{2}\right) $
we find

\begin{equation}
\Gamma _{R\uparrow }^{\left( 2\right) }\left( \mathbf{p},p_{0};\omega
\right) =p_{0}+k_{F}\left( \omega \right) \left( p_{y}+k_{F}\left( \omega
\right) +\frac{\Delta ^{2}\left( \omega \right) }{2k_{F}\left( \omega
\right) }\right) -\Sigma _{R\uparrow }\left( \mathbf{p},p_{0};\omega \right)
,
\end{equation}
where the renormalized self-energy $\Sigma _{R\uparrow }$ is determined by
the Feynman diagrams shown in Figure $2$.

The last two diagrams of $-i\Sigma _{R\uparrow }$ are originated by the
counterterms of the renormalized lagrangian and will be defined in detail
later on. Our convention here is to associate the crossed dot with the bare
coupling function $\Delta U_{0}\left( \mathbf{p}_{1},\mathbf{p}_{4}\right) $
given by

\begin{equation*}
\Delta U_{0}\left( \mathbf{p}_{1},\mathbf{p}_{4}\right) =Z\left( \mathbf{p}%
_{1}\right) Z\left( \mathbf{p}_{4}\right) U_{0}\left( \mathbf{p}_{1},\mathbf{%
p}_{4}\right) -U
\end{equation*}

\begin{equation*}
\cong \frac{4\lambda \left( \omega \right) }{2\pi ^{2}k_{F}\left( \omega
\right) }U^{2}\ln \left( \frac{\Omega }{\omega }\right) \delta _{\mathbf{p}%
_{1}+\mathbf{p}_{2}=\mathbf{0}}+
\end{equation*}

\begin{equation}
-\frac{\lambda \left( \omega \right) -\Delta \left( \omega \right) }{2\pi
^{2}k_{F}\left( \omega \right) }U^{2}\ln \left( \frac{\Omega }{\omega }%
\right) \delta _{\mathbf{p}_{1}=\mathbf{p}_{3}}\delta _{\mathbf{p}_{1}=%
\mathbf{p}_{4}=\left( 0,-2k_{F}\left( \omega \right) +\frac{\Delta
^{2}\left( \omega \right) }{k_{F}\left( \omega \right) }\right) }+...
\end{equation}

As a result of this it turns out that diagram $\left( d\right) $ reduces to

\begin{equation}
\Sigma _{R\uparrow }^{\left( d\right) }\left( \mathbf{p},p_{0};\omega
\right) =\frac{U^{2}\left( \omega \right) }{2\pi ^{4}}\lambda \left( \omega
\right) \left( \lambda \left( \omega \right) -\Delta \left( \omega \right)
\right) \left( \frac{3\lambda \left( \omega \right) +\Delta \left( \omega
\right) }{k_{F}\left( \omega \right) }\right) \ln \left( \frac{\Omega }{%
\omega }\right) +...
\end{equation}%
Since diagrams $\left( a\right) $ and $\left( b\right) $ produce\cite{Ferraz}

\begin{center}
\begin{equation*}
\Sigma _{R\uparrow }^{\left( a+b\right) }\left( \mathbf{p},p_{0};\omega
\right) =\frac{2\lambda ^{2}\left( \omega \right) }{\pi ^{2}}U-\frac{3U^{2}}{%
64\pi ^{4}}\left( \frac{\lambda \left( \omega \right) -\Delta \left( \omega
\right) }{k_{F}\left( \omega \right) }\right) ^{2}\left( p_{0}+k_{F}\left(
\omega \right) \left( p_{y}+\right. \right. 
\end{equation*}

\begin{equation*}
\left. +k_{F}\left( \omega \right) +\frac{\Delta ^{2}\left( \omega \right) }{%
2k_{F}\left( \omega \right) }\right) \cdot 
\end{equation*}

\begin{equation*}
\cdot \left[ \ln \left( \frac{\Omega +p_{0}-i\delta }{-k_{F}\left( \omega
\right) \left( p_{y}+k_{F}\left( \omega \right) +\frac{\Delta ^{2}\left(
\omega \right) }{2k_{F}\left( \omega \right) }\right) +p_{0}-i\delta }%
\right) \right. 
\end{equation*}
\end{center}

\begin{equation}
\left. +\ln \left( \frac{\Omega -p_{0}-i\delta }{k_{F}\left( \omega \right)
\left( p_{y}+k_{F}\left( \omega \right) +\frac{\Delta ^{2}\left( \omega
\right) }{2k_{F}\left( \omega \right) }\right) -p_{0}-i\delta }\right) %
\right] +...
\end{equation}%
and since diagram $\left( c\right) $ gives simply

\begin{equation}
\Sigma _{R\uparrow }^{\left( c\right) }\left( \mathbf{p},p_{0};\omega
\right) =-\Delta Z\left( \frac{\overline{p}}{\omega };\omega \right) \left(
p_{0}-\frac{p^{2}}{2}\right) -\Delta \mu _{0}\left( \frac{\overline{p}}{%
\omega };\omega \right) ,
\end{equation}%
with $\overline{p}\left( \omega \right) =k_{F}\left( \omega \right) \left(
p_{y}+k_{F}\left( \omega \right) +\frac{\Delta ^{2}\left( \omega \right) }{%
2k_{F}\left( \omega \right) }\right) $ the renormalized $\Gamma _{R\uparrow
}^{\left( 2\right) }\left( \mathbf{p},p_{0};\omega \right) $ in the vicinity
of $\mathbf{p}^{\ast }=\left( \Delta \left( \omega \right) ,-k_{F}\left(
\omega \right) +\frac{\Delta ^{2}\left( \omega \right) }{2k_{F}\left( \omega
\right) }\right) $ becomes

\begin{equation*}
\Gamma _{R\uparrow }^{\left( 2\right) }\left( \mathbf{p},p_{0};\omega
\right) =\left( p_{0}+k_{F}\left( \omega \right) \left( p_{y}+k_{F}\left(
\omega \right) +\frac{\Delta ^{2}\left( \omega \right) }{2k_{F}\left( \omega
\right) }\right) \right) \left[ 1+\right. 
\end{equation*}

\begin{equation*}
\frac{3U^{2}}{64\pi ^{4}}\left( \frac{\lambda \left( \omega \right) -\Delta
\left( \omega \right) }{k_{F}\left( \omega \right) }\right) ^{2}\left( \ln
\left( \frac{\Omega -p_{0}-i\delta }{k_{F}\left( \omega \right) \left(
p_{y}+k_{F}\left( \omega \right) +\frac{\Delta ^{2}\left( \omega \right) }{%
2k_{F}\left( \omega \right) }\right) -p_{0}-i\delta }\right) \right. 
\end{equation*}

\begin{equation*}
\left. +\ln \left( \frac{\Omega +p_{0}-i\delta }{-k_{F}\left( \omega \right)
\left( p_{y}+k_{F}\left( \omega \right) +\frac{\Delta ^{2}\left( \omega
\right) }{2k_{F}\left( \omega \right) }\right) +p_{0}-i\delta }\right)
\right) ]-\frac{2\lambda ^{2}\left( \omega \right) }{\pi ^{2}}U
\end{equation*}

\begin{equation*}
+\Delta Z\left( \frac{\overline{p}}{\omega };\omega \right) \left( p_{0}-%
\frac{p^{2}}{2}\right) +\Delta \mu _{0}\left( \frac{\overline{p}}{\omega }%
;\omega \right) 
\end{equation*}

\begin{equation}
-\frac{U^{2}\left( \omega \right) }{2\pi ^{4}}\lambda \left( \omega \right)
\left( \lambda \left( \omega \right) -\Delta \left( \omega \right) \right)
\left( \frac{3\lambda \left( \omega \right) +\Delta \left( \omega \right) }{%
k_{F}\left( \omega \right) }\right) \ln \left( \frac{\Omega }{\omega }%
\right) +...
\end{equation}

To determine $\Delta Z$ and $\Delta \mu _{0}$ we now define $\Gamma
_{R}^{\left( 2\right) }$ such that $\Gamma _{R\uparrow }^{\left( 2\right)
}\left( \mathbf{p}\ast ,\omega ;\omega \right) =\omega $, with $\omega
\approx 0$. From this prescription we find

\begin{equation}
\Delta Z\left( \frac{\overline{p}}{\omega }=0;\omega \right) =-\frac{3U^{2}}{%
32\pi ^{4}}\left( \frac{\lambda \left( \omega \right) -\Delta \left( \omega
\right) }{k_{F}\left( \omega \right) }\right) ^{2}\ln \left( \frac{\Omega }{%
\omega }\right) +...
\end{equation}%
and, similarly,

\begin{equation*}
\mu _{0}=\frac{1}{2}k_{F}^{2}\left( \omega \right) +\frac{2\lambda
^{2}\left( \omega \right) }{\pi ^{2}}U\left( \omega \right) \left[ 1+\right. 
\end{equation*}

\begin{equation}
\left. +\frac{U\left( \omega \right) }{2\pi ^{2}}\left( \frac{\lambda \left(
\omega \right) -\Delta \left( \omega \right) }{2\lambda \left( \omega
\right) }\right) \left( \frac{3\lambda \left( \omega \right) +\Delta \left(
\omega \right) }{k_{F}\left( \omega \right) }\right) \ln \left( \frac{\Omega 
}{\omega }\right) +...\right] 
\end{equation}%
If we now replace these results back into the equation for $\Gamma
_{R\uparrow }^{\left( 2\right) }$ , we find for $\mathbf{p}\approx \mathbf{p}%
^{\ast }$ and $p_{0}\approx \omega $

\begin{equation*}
\Gamma _{R\uparrow }^{\left( 2\right) }\left( \mathbf{p},p_{0};\omega
\right) =\left( p_{0}+k_{F}\left( \omega \right) \left( p_{y}+k_{F}\left(
\omega \right) +\frac{\Delta ^{2}\left( \omega \right) }{2k_{F}\left( \omega
\right) }\right) \right) \left[ 1+\right. 
\end{equation*}

\begin{equation*}
\bigskip \frac{3U^{2}}{64\pi ^{4}}\left( \frac{\lambda \left( \omega \right)
-\Delta \left( \omega \right) }{k_{F}\left( \omega \right) }\right)
^{2}\cdot 
\end{equation*}

\begin{equation*}
\cdot \left( \ln \left( \frac{\omega }{k_{F}\left( \omega \right) \left(
p_{y}+k_{F}\left( \omega \right) +\frac{\Delta ^{2}\left( \omega \right) }{%
2k_{F}\left( \omega \right) }\right) -p_{0}-i\delta }\right) \right. 
\end{equation*}

\begin{equation}
\left. +\ln \left( \frac{\omega }{-k_{F}\left( \omega \right) \left(
p_{y}+k_{F}\left( \omega \right) +\frac{\Delta ^{2}\left( \omega \right) }{%
2k_{F}\left( \omega \right) }\right) +p_{0}-i\delta }\right) \right) +...]
\end{equation}

Consequently at $p_{0}=0$ and $\mathbf{p}\approx \mathbf{p}^{\ast }$ the
renormalized self-energy is such that

\begin{equation*}
\func{Im}\Sigma _{R\uparrow }\left( \mathbf{p},p_{0}=0;\omega \right) =-%
\frac{3U^{2}}{64\pi ^{4}}\left( \frac{\lambda \left( \omega \right) -\Delta
\left( \omega \right) }{k_{F}\left( \omega \right) }\right) ^{2}k_{F}\left(
\omega \right) \left( p_{y}+\right. 
\end{equation*}

\begin{equation}
\left. +k_{F}\left( \omega \right) +\frac{\Delta ^{2}\left( \omega \right) }{%
2k_{F}\left( \omega \right) }\right) \left[ \frac{\pi }{2}\theta \left( 
\overline{p}\right) +\theta \left( -\overline{p}\right) \right] +...
\end{equation}

Thus the imaginary part of the renormalized Green%
%TCIMACRO{\U{b4}}%
%BeginExpansion
\'{}%
%EndExpansion
s function $\func{Im}G$ changes sign at $\overline{p}\left( \omega \right) =0
$. Since from our \ previous result we also have that

\begin{equation}
\func{Re}\Sigma _{R\uparrow }\left( \overline{p}=0,p_{0}=0;U\left( \omega
\right) ;\omega \right) =0
\end{equation}
we can define a dimensionless renormalized chemical potential $\overline{\mu 
}\left( \omega \right) $ through

\begin{equation}
\mu _{0}=\omega Z_{\mu }\left( \omega \right) \overline{\mu }\left( \omega
\right) =Z_{\mu }\left( \omega \right) \frac{1}{2}k_{F}^{2}\left( \omega
\right) ,
\end{equation}%
where, using our perturbative results

\begin{equation*}
Z_{\mu }\left( \omega \right) =1+\frac{U\left( \omega \right) }{\pi ^{2}}%
\left( \frac{2\lambda \left( \omega \right) }{k_{F}\left( \omega \right) }%
\right) ^{2}\left[ 1+\right. 
\end{equation*}

\begin{equation}
\left. \frac{U\left( \omega \right) }{2\pi ^{2}}\left( \frac{\lambda \left(
\omega \right) -\Delta \left( \omega \right) }{2\lambda \left( \omega
\right) }\right) \left( \frac{3\lambda \left( \omega \right) +\Delta \left(
\omega \right) }{k_{F}\left( \omega \right) }\right) \ln \left( \frac{\Omega 
}{\omega }\right) +...\right] 
\end{equation}

To go beyond this two-loop result we can invoke the renormalization group 
\textit{(RG)}. In this way, since $\omega d\mu _{0}/d\omega =0$ it follows
immediately that

\begin{equation}
\omega \frac{d\overline{\mu }\left( \omega \right) }{d\omega }=-\left(
1+\gamma _{\mu }\right) \overline{\mu }\left( \omega \right) ,
\end{equation}%
where $\gamma _{\mu }=d\ln Z_{\mu }\left( \omega \right) /d\ln \omega $.

In general we have that $Z_{\mu }=Z_{\mu }\left( \omega ;\frac{\lambda
\left( \omega \right) }{k_{F}\left( \omega \right) };\frac{\Delta \left(
\omega \right) }{k_{F}\left( \omega \right) };U\left( \omega \right) \right) 
$. To perform a full calculation of $\gamma _{\mu }$ is a difficult problem.
However \ if we assume that the renormalized coupling and the ratios of the
Fermi surface parameters are fixed point values we can use our perturbative
results to find

\begin{equation}
\gamma _{\mu }=-\frac{U^{\ast 2}}{\pi ^{4}}\left( \frac{\lambda -\Delta }{%
k_{F}}\right) ^{\ast }\left( \frac{\lambda }{k_{F}}\right) ^{\ast }\left( 
\frac{3\lambda +\Delta }{k_{F}}\right) ^{\ast }=-\gamma _{\mu }^{\ast }
\end{equation}
If we then replace this into the $RG$ equation for $\overline{\mu }\left(
\omega \right) $ we find

\begin{equation}
\overline{\mu }\left( \omega \right) =\overline{\mu }\left( \Omega \right)
\left( \frac{\omega }{\Omega }\right) ^{-\left( 1-\gamma _{\mu }^{\ast
}\right) }
\end{equation}
Hence the renormalized chemical potential parameter $\mu \left( \omega
\right) =\frac{1}{2}k_{F}^{2}\left( \omega \right) \sim \Omega \left( \frac{%
\omega }{\Omega }\right) ^{\gamma _{\mu }^{\ast }}$ , for a non-trivial
fixed point $U^{\ast }$ , is nullified by the anomalous dimension $\gamma
_{\mu }^{\ast }$ when $\omega \rightarrow 0$. Note that even if $\mu \left(
\omega \right) \rightarrow 0$ the physical potential which we define more
generally as $\mu _{R}\left( \frac{\mu }{\omega };U\left( \omega \right)
\right) $ might not be.To test this assumption we use the ansatz\cite{Gross}

\begin{equation}
\mu _{R}\left( \frac{\mu }{\omega };U\left( \omega \right) \right) =\mu
f\left( \exp t;U\left( \omega \right) \right)
\end{equation}
with $t=\ln \left( \frac{\mu }{\omega }\right) $. Since $\mu _{R}\left( 
\frac{\mu }{\omega };U\left( \omega \right) \right) $ satisfies the $RG$
equation

\begin{equation}
\left[ \omega \frac{\partial }{\partial \omega }+\beta \left( U\right) \frac{%
\partial }{\partial U}+\omega \frac{\partial \ln \mu \left( \omega \right) }{%
\partial \omega }\mu \frac{\partial }{\partial \mu }\right] \mu _{R}\left( 
\frac{\mu }{\omega };U\left( \omega \right) \right) =0
\end{equation}
it then follows that the function $f$ is determined by

\begin{equation}
\left[ -\frac{\partial }{\partial t}+\widetilde{\beta }\left( U\right) \frac{%
\partial }{\partial U}+\widetilde{\gamma }_{\mu }\left( U\right) \right]
f\left( \exp t;U\right) =0
\end{equation}
where

\begin{equation}
\widetilde{\beta }\left( U\right) =\frac{\beta \left( U\right) }{1-\omega 
\frac{\partial \ln \mu \left( \omega \right) }{\partial \omega }}
\end{equation}
and

\begin{equation}
\widetilde{\gamma }_{\mu }\left( U\right) =\frac{\omega \frac{\partial \ln
\mu \left( \omega \right) }{\partial \omega }}{1-\omega \frac{\partial \ln
\mu \left( \omega \right) }{\partial \omega }}
\end{equation}
The general solution for $f$ is of the form

\begin{equation}
f\left( \exp t;U\right) =f\left( 1;U\left( t;U\right) \right) \exp
\int_{0}^{t}d\overline{t}\widetilde{\gamma }_{\mu }\left( U\left( \overline{t%
};U\right) \right)
\end{equation}
where

\begin{equation}
\frac{\partial U\left( t;U\right) }{\partial t}=\widetilde{\beta }\left(
U\left( t;U\right) \right) ,
\end{equation}
with $U\left( t=0;U\right) =0$. If we consider the limit $\mu \rightarrow 0$
or $t\rightarrow -\infty $, assuming again that the physical system is
brought to a critical condition with $U\rightarrow U^{\ast }$, the physical
chemical potential reduces to

\begin{equation}
\lim_{\mu \rightarrow 0}\mu _{R}\left( \frac{\mu }{\omega };U\left( \omega
\right) \right) =\mu f\left( 1;U^{\ast }\right) \left( \frac{\mu }{\omega }%
\right) ^{\frac{^{\gamma _{\mu }^{\ast }}}{1-\gamma _{\mu }^{\ast }}}\sim
\mu ^{\frac{1}{1-\gamma _{\mu }^{\ast }}}
\end{equation}
Therefore only if $1>\gamma _{\mu }^{\ast }$ , $\mu _{R}\left( \frac{\mu }{%
\omega };U^{\ast }\right) \rightarrow 0$ when $\mu \rightarrow 0$. Note that
the existence of a non-trivial infrared$\left( IR\right) $ stable $U^{\ast }$
is a necessary condition for a non-zero $\mu _{R}$. Since we already proved
the existence of such fixed points elsewhere\cite{Ferraz} a non-zero $\mu
_{R}$ is clearly attainable for appropriate renormalized $FS$ parameters.
Moreover if we use our earlier estimate $U^{\ast }=\frac{8\pi ^{2}}{3}\left( 
\frac{k_{F}}{\lambda -\Delta }\right) ^{\ast }$ , the latter inequality
becomes $\left( \frac{k_{F}}{3\lambda +\Delta }\right) ^{\ast }>\frac{64}{9}%
\left( \frac{\lambda }{\lambda -\Delta }\right) ^{\ast }.$ Thus the $FS$
will suffer a further truncation in $\mathbf{k}$-space wherever this
inequality is satisfied. This result is very suggestive in view of the
observation that in the pseudogap phase of the underdoped high temperature
superconductors the $FS$ is indeed truncated and there should be an
insulating phase associated with the momentum space region where the charge
gap acquires a non-zero value. That insulating state might well be the spin
liquid proposed by Furukawa, Rice and Salmhofer earlier on\cite{Maurice}.

In conclusion we showed that in two spatial dimensions the Fermi surface is
renormalized by interactions. If the physical system approaches criticality
with non-trivial $IR$ coupling fixed points the renormalized physical
potential $\mu _{R}$ can either vanish or acquire a non-zero value. The
vanishing of $\mu _{R}$ is indicative that the Fermi surface can suffer a
further truncation and the resulting insulating state might well be an
insulating spin liquid phase.

\begin{acknowledgement}
I wish to thank T. Busche, L. Bartosch and P.Kopietz for very useful
discussions which motivated this work.
\end{acknowledgement}

Figure $1$: \ Truncated $2d$ Fermi Surface.

\begin{center}
\FRAME{ftbpF}{4.4477in}{4.6371in}{0pt}{}{}{fig1.eps}{\special{language
"Scientific Word";type "GRAPHIC";maintain-aspect-ratio TRUE;display
"USEDEF";valid_file "F";width 4.4477in;height 4.6371in;depth
0pt;original-width 1.0508in;original-height 1.0957in;cropleft "0";croptop
"1";cropright "1";cropbottom "0";filename '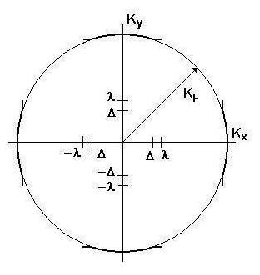';file-properties
"XNPEU";}}
\end{center}

Figure $2$: Self-energy $\Sigma _{R\uparrow }$ diagrams up to two-loops.

\begin{center}
\FRAME{ftbpF}{4.7245in}{1.599in}{0pt}{}{}{fig2.eps}{\special{language
"Scientific Word";type "GRAPHIC";maintain-aspect-ratio TRUE;display
"USEDEF";valid_file "F";width 4.7245in;height 1.599in;depth
0pt;original-width 1.7919in;original-height 0.5881in;cropleft "0";croptop
"1";cropright "1";cropbottom "0";filename '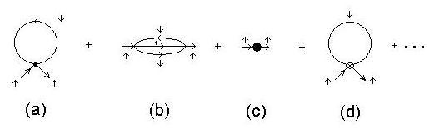';file-properties
"XNPEU";}}
\end{center}

\end{document}